\documentclass{aa}  

\usepackage[version=4]{mhchem}
\usepackage[varg]{txfonts}
\usepackage{graphicx}
\usepackage{makecell} 
\usepackage{siunitx}
\usepackage{amssymb}
\usepackage[colorlinks=true,citecolor=blue,urlcolor=blue,linkcolor=blue]{hyperref}
\usepackage{url}
\usepackage{natbib}
\usepackage{color}
\definecolor{linkblue}{rgb}{0,0,0.8}
\definecolor{linkgreen}{rgb}{0,0.5,0}

\bibpunct{(}{)}{;}{a}{}{,}

\begin{document}
\makeatletter
\@ifundefined{linenumbers}{}{%
  \let\linenumbers\relax
  \let\nolinenumbers\relax
}
\makeatother
\title{Learning cosmology from nearest neighbour statistics}

   \author{Atrideb Chatterjee\inst{1}\fnmsep\thanks{\email{a.chatterjee@rug.nl}}
         \and Arka Banerjee\inst{2}
         \and Francisco Villaescusa-Navarro\inst{3,4}
         \and Tom Abel\inst{5,6,7}
        }

   \institute{Kapteyn Astronomical Institute, University of Groningen, PO Box 800, 9700 AV Groningen, The Netherlands
   \and Department of Physics, Indian Institute of Science Education and Research, Pune 411008, India
   \and Center for Computational Astrophysics, Flatiron Institute, 162 5th Avenue, New York, NY, 10010, USA
   \and Department of Astrophysical Sciences, Princeton University, 4 Ivy Lane, Princeton, NJ 08544 USA
   \and Stanford University, Department of Physics, 382 Via Pueblo Mall, Stanford, CA 94305, USA
   \and Kavli Institute for Particle Astrophysics \& Cosmology, Stanford University, PO Box 2450, Stanford, CA 94305, USA
   \and SLAC National Accelerator Laboratory, Menlo Park, CA 94025, USA          }
 
  \abstract
   {Extracting cosmological parameters from galaxy and halo catalogues to sub-per cent level accuracy is an important aspect of modern cosmology, especially in view of ongoing and upcoming surveys such as Euclid, DESI, and LSST. While traditional two-point statistics have been known to be suboptimal for this task, recently proposed k-nearest neighbour (kNN) based summary statistics have demonstrated a tighter constraining power. Building on the kNN statistics, we introduced a new field-level representation of discrete halo catalogues:  NN distance maps. We employed this technique on the halo catalogues obtained from Quijote N-body simulation suites. By combining these maps with kNN-based summary statistics, we trained a hybrid neural network to infer cosmological parameters, showing that the resulting constraints achieve state-of-the-art accuracy, comparable to the best existing methods. In addition, our hybrid framework is $5-10$ times more computationally efficient than some of the existing point-cloud-based ML methods.}


   \maketitle

\section{Introduction}
\label{sec:intro}
Within the standard model of cosmology, Lambda cold dark matter ($\Lambda$CDM) and its extensions, values of the cosmological parameters encode the fundamental properties of our Universe, ranging from the matter-energy content to the expansion history. Therefore, achieving sub-per cent level precision in measuring these parameters is essential for a detailed understanding of the origin and evolution of the Universe. To this end, the large-scale structure (LSS) of the Universe — the cosmic web traced by galaxies and halos — provides a rich and direct observable imprint of these parameters, making it a powerful probe for constraining cosmological parameters \citep[see {e.g.}][for recent studies]{EFTLSS_damico, EFTLSS_colas, Uhlemann_2020, Villaescusa_Navarro_2020, Gualdi_2021, Valgiannis_2021, Liu_2022, Ajani_2020}. This is one of the primary science objectives for many new and upcoming surveys, for example CMB-S4 \citep{CMBS4_22}, the Roman Space Telescope \citep{Roman_23}, the Dark Energy Spectroscopic Instrument (DESI) \citep{DESI}, Euclid \citep{Laureijs2011, Euclid2016, Euclid2022-Tiago_Castro}, the Large Synoptic Survey Telescope (LSST) \citep{Ivezic_19}, and J--PAS \citep{Benitez2014}. 

To extract cosmological information from LSS data, traditional approaches have relied on summary statistics, usually the two-point  correlation function or its Fourier space equivalent, the power spectrum $P(k)$. While the two-point correlation function is computationally efficient and well-established, it captures, by definition, only the Gaussian aspect of clustering of matter or tracers. This makes the statistic suboptimal and inadequate on non-linear scales where the density field develops non-Gaussian features. In light of this, multiple approaches have been developed to go beyond the two-point function \citep[see  {e.g.}][]{Marques_2019, Villaescusa_Navarro_2020, Gualdi_2020, Hahn_2020, Friedrich_2020, Giri_2020, Harnois_2021a, Samushia_2021, Naidoo_2021, Bayer_2021, Eickenberg_22}. While higher-order statistics can, in principle, capture more of the information from the underlying fields, they often come at a significant computational cost. These considerations will play an important role in the deployment of these techniques to the vast datasets that will soon be available in cosmology. 

In parallel, recent advances in machine learning (ML) have opened new avenues to analyse LSS data by leveraging neural networks to learn complex, high-dimensional features \citep{Siamak_16, Fluri_19, Makinen_graph_22, Niall_2020, Gillet_19, Hortua_2021, Pablo_Graph, GNN_MW_M31, Hasan_HIFLOW, Pointnet_22, Cosmology_with_One_Galaxy_Villaescusa-Navarro:2022twv, Cuesta_pointcloud_23, Domain_adaptive_roncoli, Ho_SBI_24, Lee_25}. This represents an alternative route to the summary statistics in terms of capturing the total information in cosmological fields, thereby constraining the values of the parameters of interest. However, standard image-based ML approaches such as convolutional neural networks (CNNs) have been demonstrated to be sub-optimal for large but spatially sparse datasets. They suffer from information loss when mapping sparse galaxy or halo catalogues to dense grids. On the other hand, point cloud methods, including graph neural networks (GNNs) \citep{desanti2023, Cuesta_pointcloud_23, shao_et_al, Makinen_graph_22, Lee_25} and PointMLP variants \citep{Pointnet_22, Chatterjee_25}, although directly applied to discrete data, face severe computational and memory constraints when applied to large galaxy catalogues with $\sim 10^5$ objects.

For this paper we formulated a new field-level representation of discrete point datasets that converts them to continuous spatial maps by assigning to every point in space a value equal to the distance to the $k$-th NN in the original point dataset. This approach was inspired by the k-nearest neighbour cumulative distribution functions ($k$NN-CDFs), introduced in \cite{Banerjee_2021a}, which have emerged as a promising cosmological summary statistic, offering an efficient and robust measure on discrete data that captures higher-order spatial correlations beyond two-point statistics. We fed the nearest neighbours (NN) distance maps created from halo fields of N-body simulations run at different cosmological parameters to train a neural network and study the possible constraints on cosmological parameters. We also studied the effects of combining the NN distance maps with the CDFs, using them as joint inputs to a hybrid neural network for cosmological parameter inference and to demonstrate tighter constraints than possible with each input individually. Our method represents a promising avenue for applying ML techniques effectively to large discrete datasets from large-scale cosmological surveys.
\section{Nearest neighbour maps from simulation data}
\label{sec:NN_stats}

\cite{Banerjee_2021a} introduced the $k$-NN cumulative distribution function as a set of useful summary statistics for the clustering of discrete data points (e.g. a catalogue of halo or galaxy positions in a typical cosmological dataset). At any scale, $r$, the value of the $k$NN-CDF represents the fraction of all points (covering the entire space over which the clustering is to be measured) that contain at least $k$ data points within a sphere of radius $r$. Using data structures like search trees, the $k$NN-CDFs can be computed quickly  on $\mathcal O(N \log N)$ time. Crucially, \cite{Banerjee_2021a} demonstrated that each $k$NN-CDF is related to different combinations of all $N$-point functions of the underlying clustering, making these statistics sensitive to all orders of beyond-Gaussian clustering. This sensitivity translates to tighter constraints on cosmological parameters than the two-point function while using the same datasets, while the quick compute time allows these statistics to be easily evaluated from large datasets.

\cite{Banerjee_2021b} extended the formalism to capture cross-correlations (at all orders) between discrete datasets, while \cite{Banerjee_2022} further extended the formalism to include cross-correlations between discrete point data and continuous maps. The application of similar ideas to measure auto-clustering and cross-clustering in sets of continuous maps, specifically weak lensing mass maps from galaxy surveys, was addressed in \cite{Anbajagane_2023}. These summary statistics have already been measured in the context of various datasets \citep{Wang_2022,  Gupta_2024, Coulton_2024, Zhou_2025, Chand_2025}, while modelling these statistics as a function of cosmological parameters and different galaxy-halo connection models has also been explored \citep{Banerjee_2022_HEFT, Yuan_2023}. The $k$-NN statistics performed well in the community-wide `Beyond-two-point challenge' and were able to recover the true cosmology with tight error bars \cite{Krause2024}.

\citet{Gangopadhyay_2025} demonstrated that the $k$NN-CDFs and their derivatives have geometric interpretations. For example, the $1$NN-CDF and its first three derivatives at some scale $r$ encode the geometry of intersections of spheres of radius $r$ centred on the data points. Specifically, the value of the CDF is proportional to the volume enclosed; the first derivative is proportional to the exposed area; the second derivative contains information about the angles of intersections of the spheres; and finally, the third derivative is related to the Euler characteristic of the resultant configuration. These connections also relate the $k$NN-CDFs geometrically to the germ-grain Minkowski Functionals \cite[see {e.g.}][]{Schmalzing_1995}. However, while the $k$NN-CDFs capture a great deal of the information about the clustering of a set of points, it is important to ask whether they can capture all the  information, and if they do not, how the missing information can be accessed. A simple way to see that the $k$NN-CDFs may not contain all clustering information is to realise that the construction of the CDF, by definition, throws away spatial information. That is, the value of the CDF at some scale $r$ retains no information about the spatial distribution of points contributing to that value. The question arises of whether they come from specific locations in the volume, or if they are   close to being uniformly distributed over the entire volume.

This motivates the construction and analysis of  the NN distance maps, where each point in space is assigned a value equal to the distance from that point to the $k$-th NN data point. This results in a smooth, continuous field-level representation of the original set of discrete data points. These maps, in their raw form, preserve spatial information about how fast or slowly the NN distances change in different directions, which, in turn, is directly related to the exact configuration of the original data points. In fact, for 1NN distance maps in three dimensions, the points where the field has a maximum in one direction lie on the face of the Voronoi tessellation defined by the data points. That is, they are equidistant from two data points. Points with maxima along two directions define the Voronoi edges and are equidistant from three data points. Finally, maxima in this field, where the gradient vanishes, correspond to Voronoi nodes, or points that are equidistant from four data points. This is demonstrated in two-dimensional slices in the left panel of Fig. \ref{fig:NN_map}. Similarly, one can construct maps for larger $k$;  the right-hand panel of Fig. \ref{fig:NN_map} shows the 2D maps for $k=4$. These ideas, especially for the first NN distances, have been explored in the field of image processing, and are referred to as the distance transform \cite[see {e.g.}][]{Jones_2006}. 

Neural networks are particularly well-suited to analysing the information in these maps. The traditional failing of CNNs and related techniques for sparse data can now be circumvented, as the mapping converts the data to a densely sampled continuous field. For this study, we applied kNN statistics (CDFs as well as maps) on the halo catalogues obtained from the Quijote simulation suite.\footnote{\url{https://quijote-simulations.readthedocs.io/en/latest/}} In particular, we focused on the 2000 Latin Hypercube realisations, each containing $512^3$ dark matter simulation particles at z=0. The cosmological parameters of these simulations vary in the range
\begin{eqnarray*}
0.1 \leq &\Omega_{\rm m}& \leq 0.5,\\
0.03 \leq &\Omega_{\rm b}& \leq 0.07,\\
0.5 \leq &h& \leq 0.9,\\
0.8 \leq &n_s& \leq 1.2,\\
0.6 \leq &\sigma_8& \leq 1.0.
\end{eqnarray*}

In this work, we   concentrate on just two of these parameters: $\Omega_{m}$ and $\sigma_8$. Further, we select $10^5$ most massive halos\footnote{The halos are identified using an FoF algorithm from the simulation particles} from the halo catalogues from each of these simulations and compute the NN distance maps and CDFs as mentioned below.\footnote{We have done a proof-of-concept calculation for maps and CDFs corresponding to the dark matter particles as well in the Appendix \ref{sec:DM_constraints}.} This number cut translates to a mass cut of around $5 \times 10^{13} \rm h^{-1}M_{\odot}$ \citep{Banerjee_2021b}, i.e. about five times more massive than   the least massive halos. This choice ensures that our result should not depend on the resolution of the
simulations.

\subsection{Generation of NN maps}
\label{subsec:NN_maps}
\begin{figure*}
    \centering
    \includegraphics[width=\textwidth]{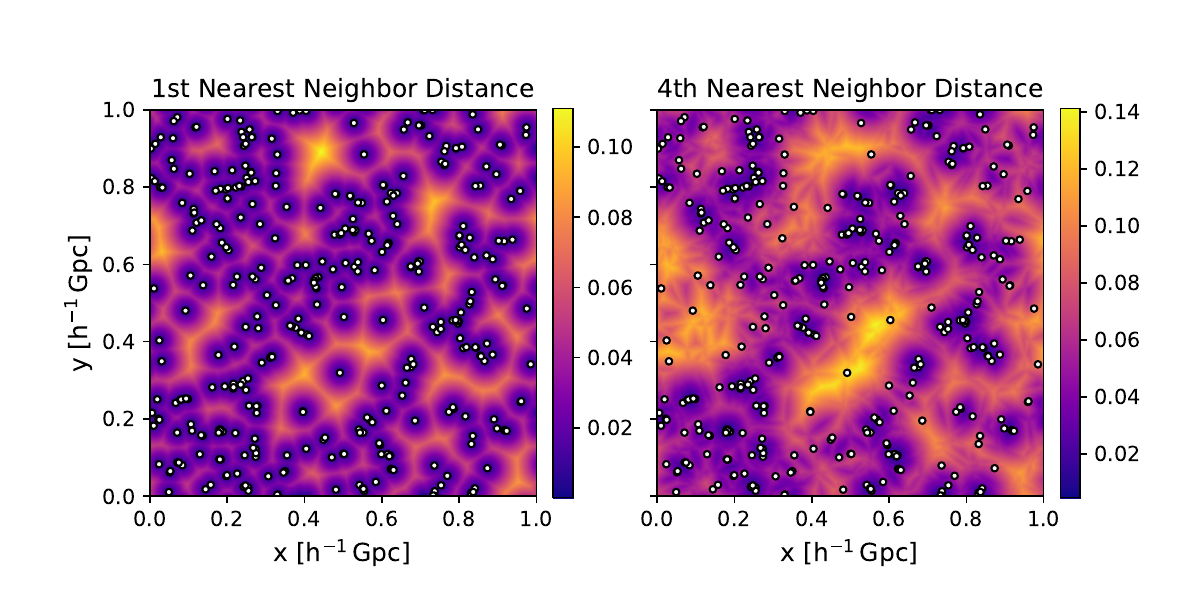}
    \caption{2D slice of the first (left) and fourth (right) nearest neighbour distance maps for one of the simulations in the Quijote simulation suites used in this study. Each pixel is coloured by the distance from the pixel to the nearest data point (left panel) and by the distance to the fourth nearest neighbour data point (right panel). As can be seen, this converts the discrete dataset into a smooth, continuous map. The colour bar represents the distance (in Gpc/$h$) from the halos. We note that these maps are only for the purpose of visualisation. They were produced with $256^2$ random query points in a $256 \times 256$ 2D grid, whereas the actual maps used in this study were produced with $10^4$ random query points in a $100 \times 100$ 2D grid, as mentioned in \ref{subsec:NN_maps}.}
    \label{fig:NN_map}
\end{figure*}

For creating the maps, we start with the 3D positions of the $10^5$ most massive halos from each simulation:

\begin{enumerate}
    \item We first divide the 3D data into 100 2D slices along the z-axis with an inter-plane distance of 10 Mpc.    
    \item For each of these slices, we generate $10^4$ query points in a $100 \times 100$ regular grid\footnote{We have also checked our result with a $256 \times 256$ grid with $256^2$ query points, but did not find any significant improvement for the Map-only scenario. We then used the $100 \times 100$ grid with $10^4$ query points for NN maps throughout the analysis.} and calculate distances of first, second, third, and fourth NNs from those query points.\footnote{As shown in \cite{Banerjee_2021b}, most of the information is contained in 1NN, 2NN, 3NN, and 4NN CDF statistics. Keeping this in mind, we started with these four NNs, but then also included 8th, 16th, and even 64th NN in the analysis. We did not find any significant change in the results, and therefore reported the constraints obtained using 1NN, 2NN, 3NN, and 4NN statistics.} 
    \item Once obtained, these distances are used to create 2D maps of kNN distances for the query points. Figure \ref{fig:NN_map} shows an example of the first and fourth NN distance maps, with the positions of halos in that slice overlaid as white points. We note that the maps shown in Fig. \ref{fig:NN_map} are produced with $256^2$ random query points in a $256 \times 256$ grid and are made only for the purpose of visualisation.
    \item Finally, for each simulation, we randomly select ten maps corresponding to each NN distance to be used in the subsequent analysis. As we are taking ten random slices from each simulation, we have a total of $20,000$ maps.  
\end{enumerate}

\subsection{Generation of the NN CDFs}
\label{subsec:NN_CDF}

\begin{figure*}
\sidecaption
  \includegraphics[width=12cm]{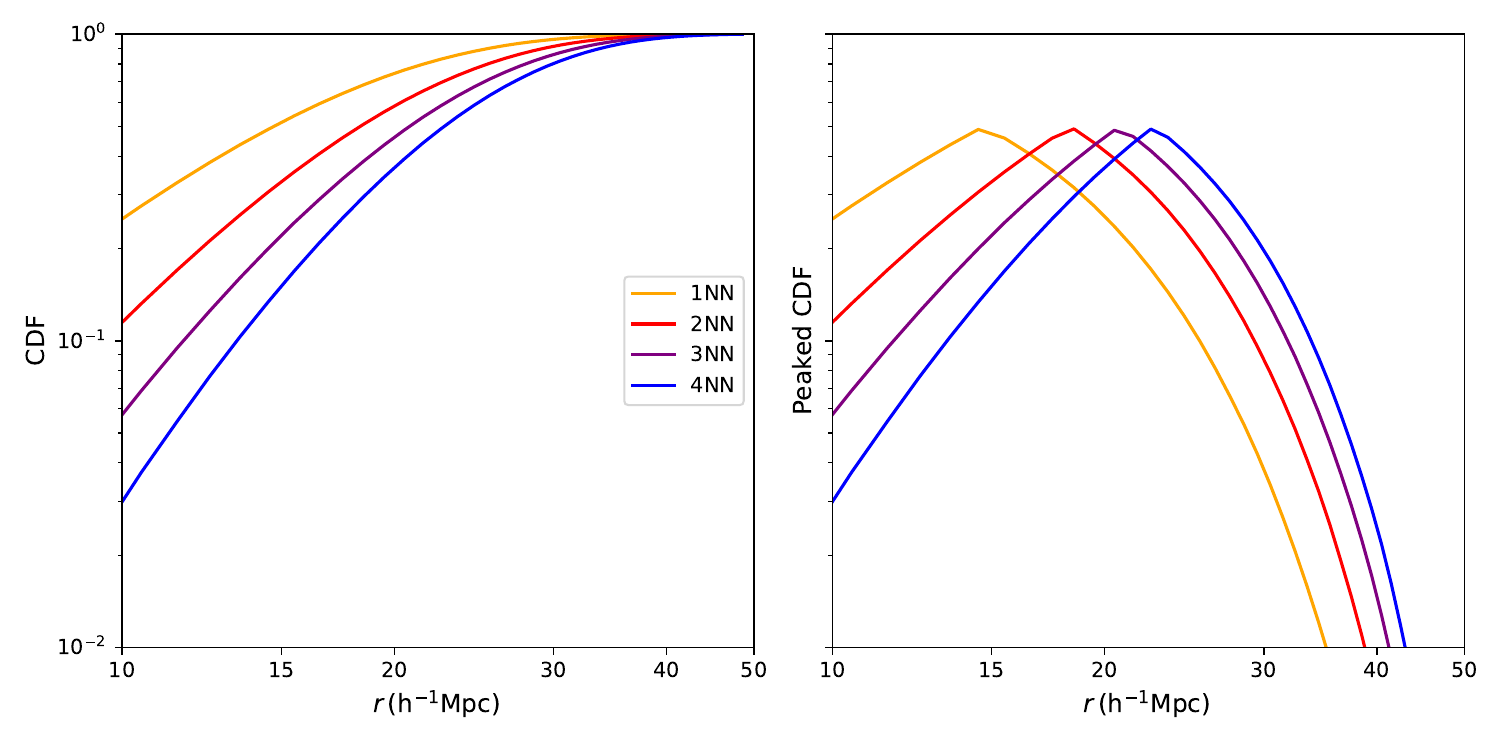}
     \caption{CDF (left panel) and peaked CDF (right panel) for 1NN (orange), 2NN (red), 3NN (magenta), and 4NN (blue) corresponding to one of the Quijote simulations in this study.}
     \label{fig:NN_CDF}
\end{figure*}

To obtain NN CDFs from the downsampled halo catalogues of the $10^5$ most massive halos, we used the following method:
\begin{enumerate}
\item We generate $10^6$ random query points in 3D within the simulation box. By taking $10^6$ random query points, we ensure that the number of query points is more than the discrete data points. 

\item Using scipy.spatial.KDTree, we compute the distances to the first, second, third, and fourth NNs for these query points. 

\item We then construct cumulative distribution functions (CDFs) of the NN distances, binned into 50 intervals of width $1\ \mathrm{Mpc}\ h^{-1}$. In the left panel of Fig.  \ref{fig:NN_CDF}, we illustrate the shapes of these functions obtained from one of our simulations.\footnote{For one of the simulations, the minimum distance for first (1NN), second (2NN), third (3NN), and fourth (4NN) nearest neighbour are $0.18, \, 0.98, \, 2.08, \, 2.94, \, 5.61 \, {\rm h^{-1} Mpc} $, respectively, where as the mean separation between two halos (after downsampling the data to $10^5$ most massive halos) is $\sim 22 \, \rm{h^{-1} Mpc}$.} For better visualisation, especially when the value of the CDFs is $\sim 1$, we depict the peaked CDF distribution for the first, second, third, and fourth NNs in the right-hand panel of the same figure. Following \cite{Banerjee_2021a}, we define the Peaked CDF as

\begin{equation}
\mathrm{PCDF}(r) =
\begin{cases}
\mathrm{CDF}(r), & \text{if } \mathrm{CDF}(r) \le 0.5, \\[4pt]
1 - \mathrm{CDF}(r), & \text{if } \mathrm{CDF}(r) > 0.5.
\end{cases}
\end{equation}
\end{enumerate}

\section{Machine learning model}
\label{sec:ML_models}
We explored three ML models: (1) Map-only: a standard ResNet model that learns from the 2D slices of NN distance maps, (2) CDF-only: a fully connected NN that learns from the NN CDFs, and (3) Map+CDF: a hybrid model that takes as input the combination of the NN CDFs and NN distance maps. We describe them below in more detail.
\subsection{Architecture}

Map-only:
We employed the \textsc{ResNet-18} \citep{ResNet} architecture, a widely used convolutional neural network designed for image-based tasks, to infer cosmological parameters from 2D NN distance maps. The main novelty of the residual network (ResNet) is in its use of skip connections that effectively mitigate the vanishing gradient problem and enable the training of very deep architectures. In our case, each input map corresponds to a 2D grid of NN distances computed from dark matter or halo slices, with separate channels for different NN distance maps (e.g. 1NN, 2NN, 3NN, 4NN). The \textsc{ResNet-18} model processes these maps through a series of convolutional, batch normalisation, ReLU, and residual blocks, ultimately reducing them to a feature vector that is passed through a fully connected layer to predict the cosmological parameters. 

CDF-only:
For each simulation, once we calculate the CDFs of the first, second, third, and fourth NN distances, each representing a 1D array with 48 bins (uniform bin width of $1\,h^{-1}$Mpc), they are then concatenated into a single 1D array of length 192 ($4 \times 48$), which serves as the input for the fully connected neural network containing multiple layers of perceptrons (MLPs). We use ReLU as the non-linear activation function in this model. The number of layers, neurons, learning rate, weight decay, and dropout are hyperparameters optimised with \textsc{OPTUNA}\citep{akiba2019optuna} in over 100 trials, as mentioned in Table \ref{tab:hyperparameters_all}.

Map+CDF:
The concept of hybrid or multimodal neural network, combination of different architecture trained on different types of data (e.g. images, texts) in order to maximise the performance of a neural network, has shown a lucrative performance in different fields of artificial intelligence \citep{Chen_23, azevedo2024hybrid, zeng2022parking, shi2019hybrid, halbouni2022cnn, siraj2020hybrid, demiss2024application, Dattilo_19} and in cosmology \citep{Dattilo_19, Ntampaka_19, Makinen_hybrid_25, Hosseini_25, Patchnet_25}. While \cite{Ntampaka_19} used the hybrid network combining the 2D images and two-point power  spectrum obtained from a simulated galaxy density field and showed that the hybrid network performs better than the individual neural network, \citet{Hosseini_25} employed  a combination of a convolutional neural networks and recurrent neural networks on 21 cm brightness temperatures and reconstructed  a 21 cm global signal with a prediction accuracy of $99.93\%$.

\begin{figure*}
   \centering
    \includegraphics[width=\textwidth]{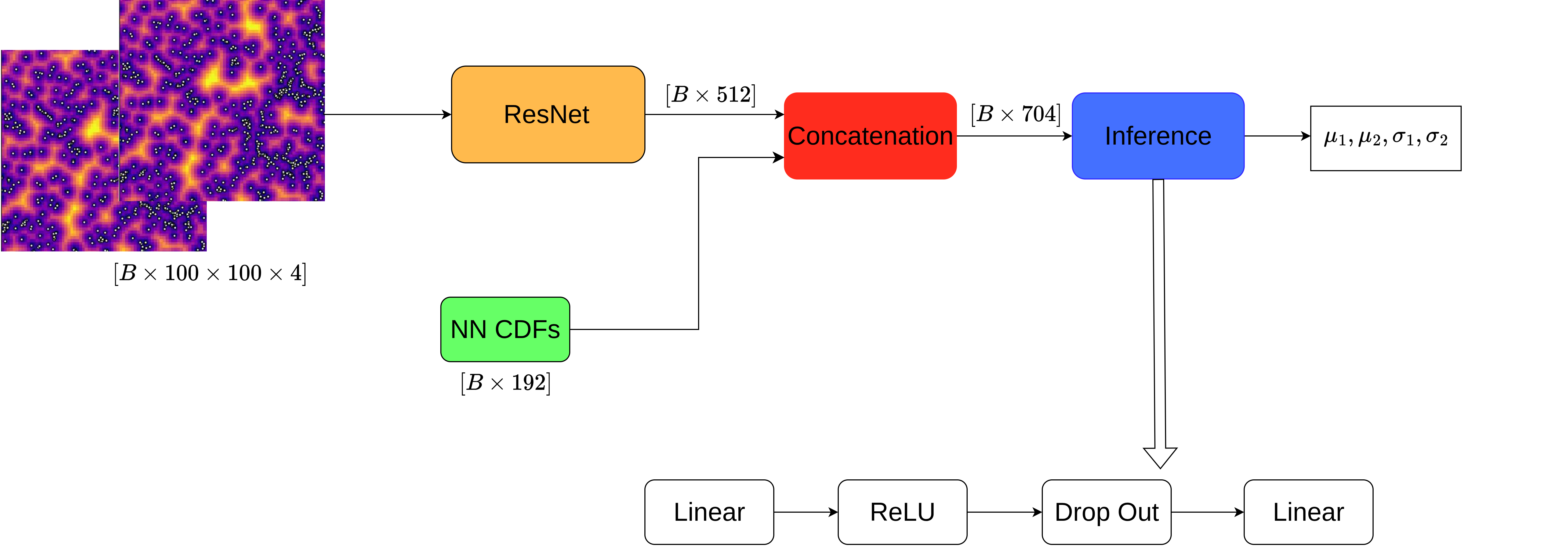}
    \caption{Hybrid network in this study. The NN distance maps are used as input to the ResNet block. The output of the ResNet is then concatenated with the NN CDFs, and the merged input then passes through the inference blocks (containing several linear, ReLU, and dropout layers) to predict the mean and standard deviation of the inferred cosmological parameters. The values in brackets show the dimension of the tensor in different stages of the architecture. Here B denotes the batch dimension.}
    \label{fig:network}
\end{figure*}

Motivated by the aforementioned works, we implemented a hybrid neural network architecture for this study. The model integrates a \textsc{ResNet-18} backbone with fully connected layers that incorporate summary statistics, i.e. the NN cumulative distribution functions (CDFs). We initialise the \textsc{ResNet-18} model after modifying the input convolutional layer to accept multiple input channels, corresponding to the different NN distances (e.g. 1NN to 4NN). The fully connected classification head of ResNet is removed and replaced with an identity layer to extract the learned feature representation from the input maps. The length of the learned features, i.e. the number of output channels from the ResNet (as shown in the figure) 
is kept to 512. These features are then concatenated with the 1D summary statistics vector (of length 192, corresponding to the concatenated CDFs from 1NN–4NN). The combined feature vector (of dimension 704, i.e. 512 from ResNet and 192 from CDFs) is passed through an inference block consisting of a multi-layer perceptron followed by a final output layer that predicts the cosmological parameters as shown in Fig. \ref{fig:network}. The number of layers and neurons in the multi-layer perceptron of the network, as well as the learning rate and weight decay, are hyperparameters optimised with \textsc{OPTUNA}\citep{akiba2019optuna} in over 100 trials, as mentioned in Table \ref{tab:hyperparameters_all}.\footnote{In the figure, the structure of the inference block presents the best architecture achieved after \textsc{OPTUNA} hyper-parameterisation.} The architecture allows the model to jointly learn from   spatial features in the NN distance maps and from global statistical summaries. 

\subsection{Loss function}  
\label{subsec:loss_func}
All the above-mentioned models are trained to perform likelihood-free inference on the value of the cosmological parameters. For each parameter, the model predicts the marginal posterior mean ($\mu$) and standard deviation ($\sigma$), defined as

\begin{eqnarray}
\mu(\mathcal{P}) &=& \int_{\theta} p(\theta | \mathcal{P}) \theta d\theta~,\\
&& \nonumber \\
\sigma^2(\mathcal{P}) &=& \int_{\theta} p(\theta | \mathcal{P}) (\theta - \mu)^2 d\theta~,
\label{Eq:mean_std}
\end{eqnarray}
where $\mathcal{P}$ represents the input, which could be CDFs for CDF-only, Maps for Map-only, or a combination of CDF and Maps for Map+CDF. To achieve this, we minimise the following loss function \citep{moment_networks, CMD} 

\begin{eqnarray}\label{eqn:loss_function}
\mathcal{L}&=& \sum_{i=1}^{N_{\theta}}\log\left(\sum_{j\in{\rm batch}}\left(\theta_{i,j} - \mu_{i,j}\right)^2\right)\nonumber \\
+&& \sum_{i=1}^{N_{\theta}} \log\left(\sum_{j\in{\rm batch}}\left(\left(\theta_{i, j} - \mu_{i,j}\right)^2 - \sigma_{i,j}^2 \right)^2\right)~,
\label{Eq:loss}
\end{eqnarray}
where $\theta_{i,j}$, $\mu_{i,j}$, and $\sigma_{i,j}$ represent the true, inferred mean, and inferred standard deviation of the parameter $i$ for the sample $j$. Further, $N_{\theta}$ is the total number of cosmological parameters; in this study there are two: $\Omega_m$ and $\sigma_8$. We employ this loss function to directly obtain the estimates of moments of the marginalised distribution of all parameters without calculating the posterior density.

\subsection{Training procedure}
We split the available input data into training (80\%), validation (10\%), and testing (10\%). For all the ML architectures, we used the Adam optimiser \citep{Kingma2014AdamAM}.\footnote{The learning rate and weight decay of the optimiser are kept as hyperparameters and later optimised using \textsc{OPTUNA}\citep{akiba2019optuna}} We used a batch size of 32 and trained for 300 epochs. Training was performed on a single NVIDIA A100 GPU and took approximately 30–120 minutes per training run, depending on the specific ML model used.

\subsection{Validation metrics}
For each cosmological parameter, we employed four statistics to quantify the accuracy and precision of our models on the test dataset, given by the following:
\begin{itemize}
\item The mean relative error, $\epsilon$, defined as 
\begin{equation}
    \epsilon = \frac{1}{N} \sum_{j = 1}^{N} \frac{|\theta_{j} - \mu_{j}|}{\theta_{j}},
\end{equation}
where N denotes the size of the test dataset. A smaller value of $\epsilon$ indicates greater precision of the network;

\item The coefficient of determination, $R^2$, defined as
\begin{equation}
    R^2 = 1 - \frac{\sum_{j = 1}^{N} (\theta_{j} - \mu_{j})^2}{\sum_{j = 1}^{N} (\theta_{j} - \overline{\theta})^2},
\end{equation}
where $\overline{\theta}=\frac{1}{N}\sum_{j= 1}^{N} \theta_{j}$. A network with higher accuracy leads to an $R^2$ value closer to 1.

\item The mean squared error, MSE, defined as

\begin{equation}
    {\rm MSE} = \frac{1}{N}\sum_{j = 1}^{N} (\theta_{j} - \mu_{j})^2.
\end{equation}
An accurate network indicates a smaller mean squared error;

\item The $\chi^2$ value, defined as 
\begin{equation}
\chi^2=\frac{1}{N}\sum_{j = 1}^{N} \frac{(\theta_{j} - \mu_{j})^2}{\sigma_{j}^2}.
\label{Eq:chi2}
\end{equation}
A value of $\chi^2$ close to 1 indicates that network errors are calibrated correctly.
\end{itemize}
\section{Results}
\begin{figure*}
    \includegraphics[width = 0.5\linewidth]{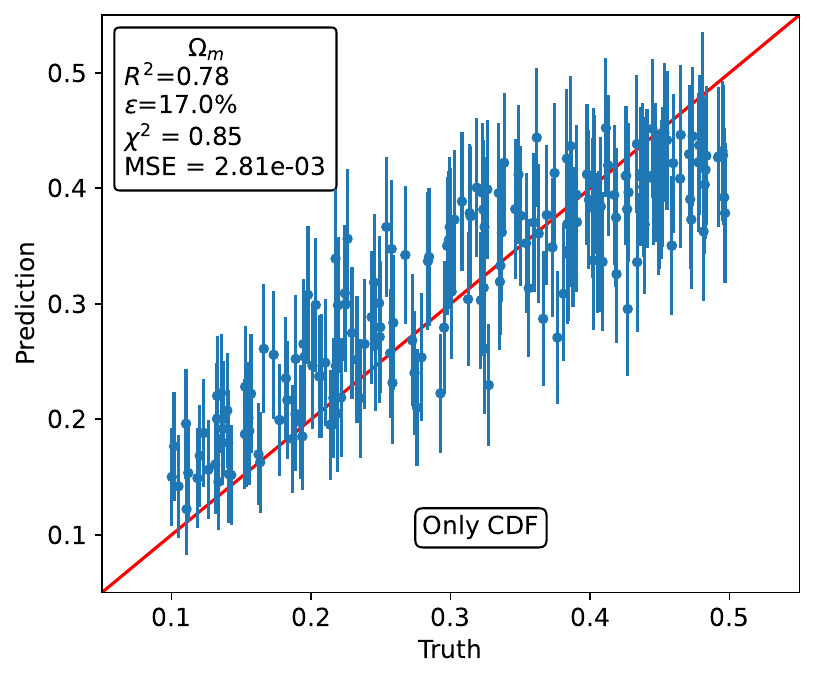}
    \includegraphics[width = 0.5\linewidth]{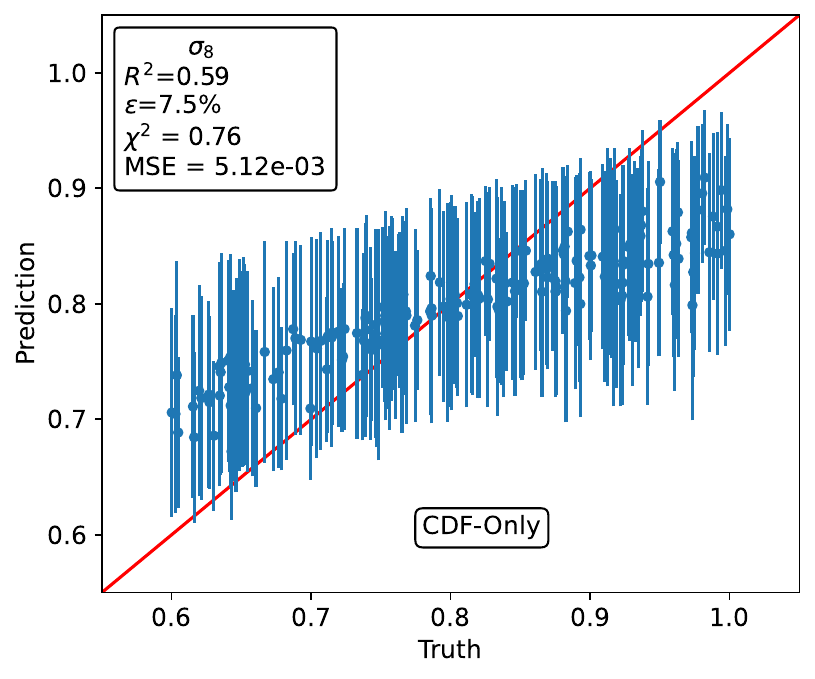}
    \includegraphics[width = 0.5\linewidth]{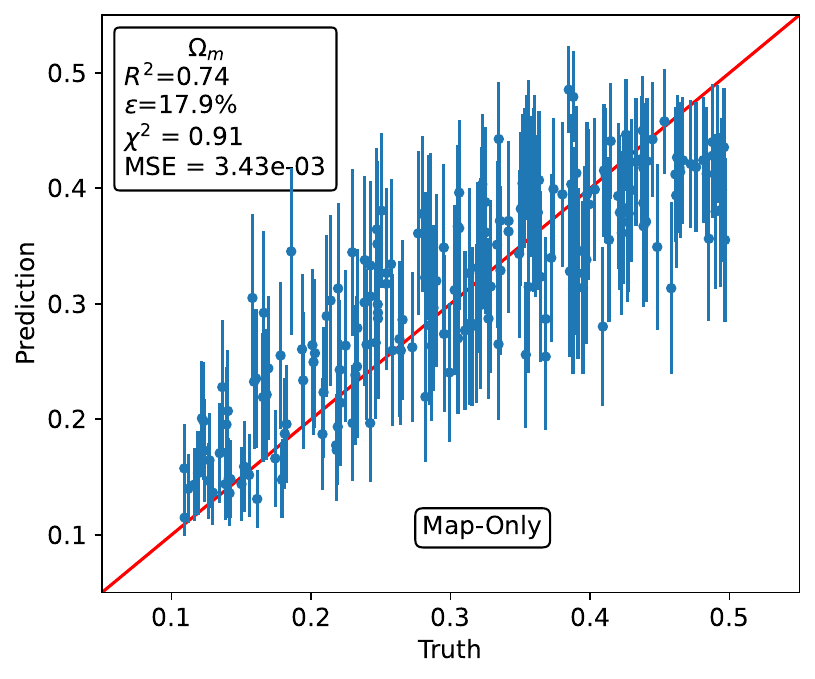}
    \includegraphics[width = 0.5\linewidth]{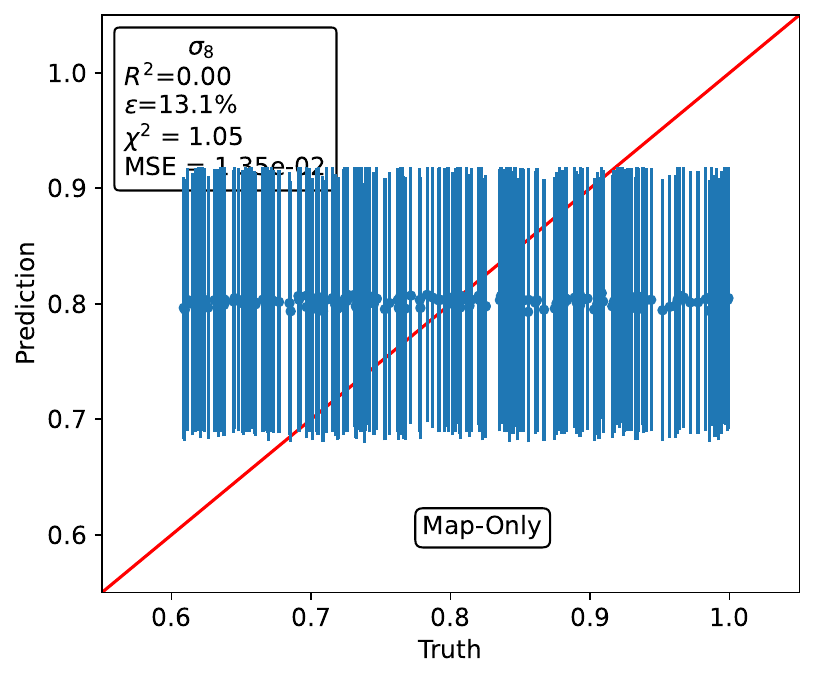}
    \includegraphics[width = 0.5\linewidth]{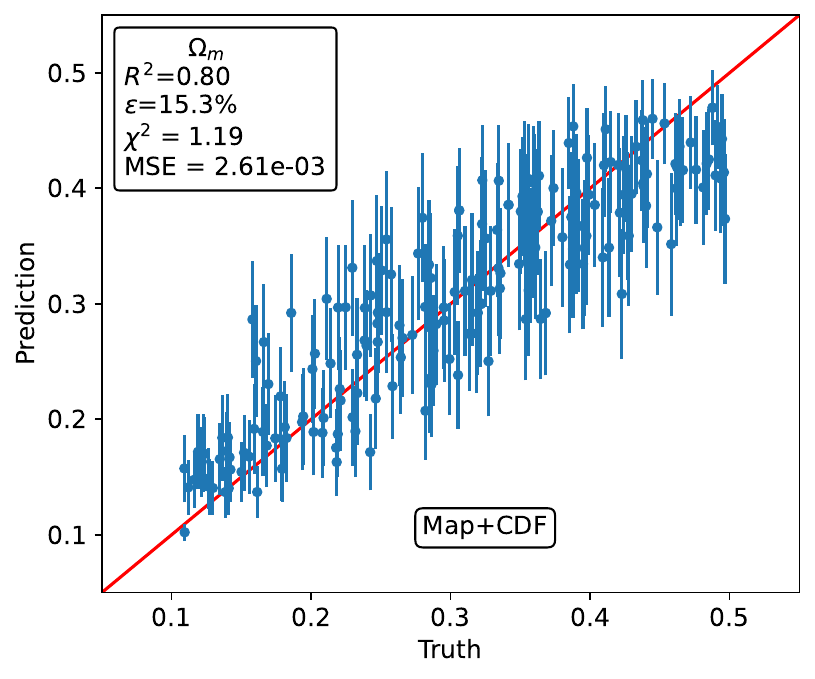}
    \includegraphics[width = 0.5\linewidth]{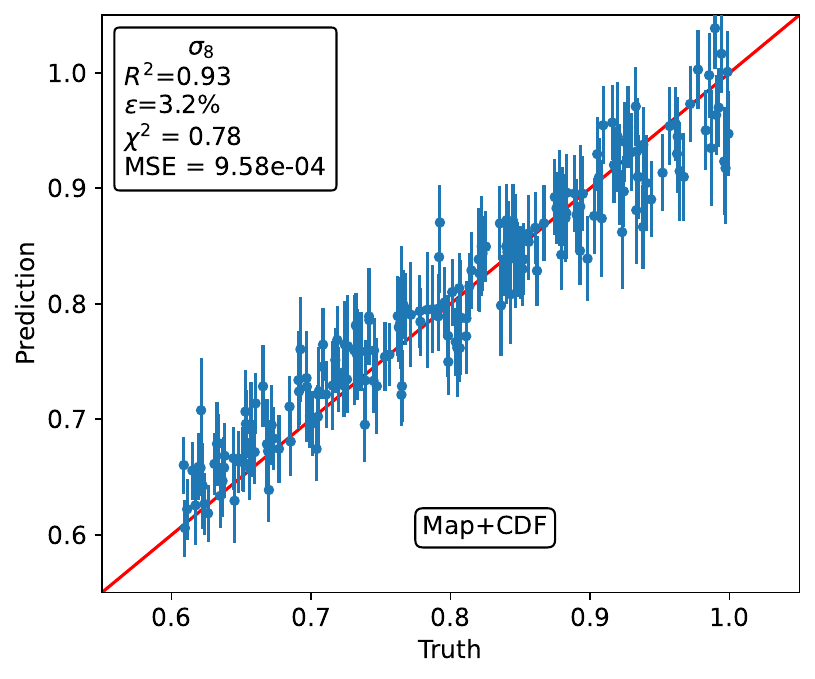}
    \caption{Performance of different models when trained to predict likelihood-free inference on   the values of $\Omega_{m}$ (left column) and $\sigma_{8}$ (right column) in three scenarios:  Top row: CDF-only;  Middle row: Map-only;  bottom row: Map+CDF. The values for different validation metrics are given in the legend. As can be seen, the Map-only scenario (middle panel)   performs worse than the CDF-only (top panel) scenario. Further, the Map+CDF model performs the best across all the validation metrics.}
    \label{fig:Om_sig8}
\end{figure*}

We now present the results obtained from the ML models studied in this work. Table \ref{tab:res_all} summarises the results for the different ML models. Figure \ref{fig:Om_sig8} shows the constraints we derive on $\Omega_{\rm m}$ and $\sigma_8$ from different ML models. We now describe the main findings in the different models:

\begin{itemize}
    \item CDF-only: From table \ref{tab:res_all}, we can see that the CDF-only scenario produces a reasonable value of relative error and $R^2$ for $\Omega_m$. The prediction for $\sigma_8$ is poor with large error bars.

    \item Map-only: When we train on NN distance maps using ResNet, we find that our models perform more poorly than the CDF-only scenario in all the validation metrics. In fact, the network fails in predicting any sensible constraints on $\sigma_8$. Although we note that the Map-only model is not a fully 3D statistic (we generate 2D slices of the NN maps as mentioned in Sect. \ref{subsec:NN_maps}), and therefore, a comparison with CDF-only should be made with caution, it is still quite contrary to our expectations. In fact, we argue in Section \ref{subsec:NN_maps} that NN distance maps would contain more information than the CDFs as they preserve the spatial information. Interestingly, we note that a very similar behaviour is seen in \cite{Chatterjee_25} and \cite{Makinen_graph_22} when used on the Qujote simulation, although it uses  a completely different network architecture. One possible explanation could be the limited number of training datasets (maps corresponding to $1600$ different parameter combinations).  It is plausible that due to the limited number of training datasets, the network is unable to break the bias-$\sigma_8$ degeneracy. When we use the Map-only method on the maps produced from DM particles (Appendix \ref{sec:DM_constraints}) rather than halos, we find that this method outperforms the CDF-only methods, as expected from an unbiased tracer. Since halo bias modifies the amplitude of large-scale statistics in a way that is degenerate with the effect of $\sigma_8$, a network trained on NN distance maps may struggle to disentangle the two for a limited amount of training data. 
    
    \item Map+CDF: The hybrid network model, where NN CDFs along with the NN distance maps are used, performs the best. It results in an $R^2$ score of $0.80$ and $0.93$ for $\Omega_m$ and $\sigma_8$, respectively. Further, we achieve a relative error of $\sim 15\%$ and $\sim 3\%$ for $\Omega_m$ and $\sigma_8$, respectively.
\end{itemize}

We note that in all the scenarios, the value of the $\chi^2$ is $ \sim 0.8$. However, this is reasonable for a parameter inference study with ML.

\subsection*{Comparison with the two-point correlation function}
\begin{figure*}
    \begin{center}
        \includegraphics[width=0.45\linewidth]{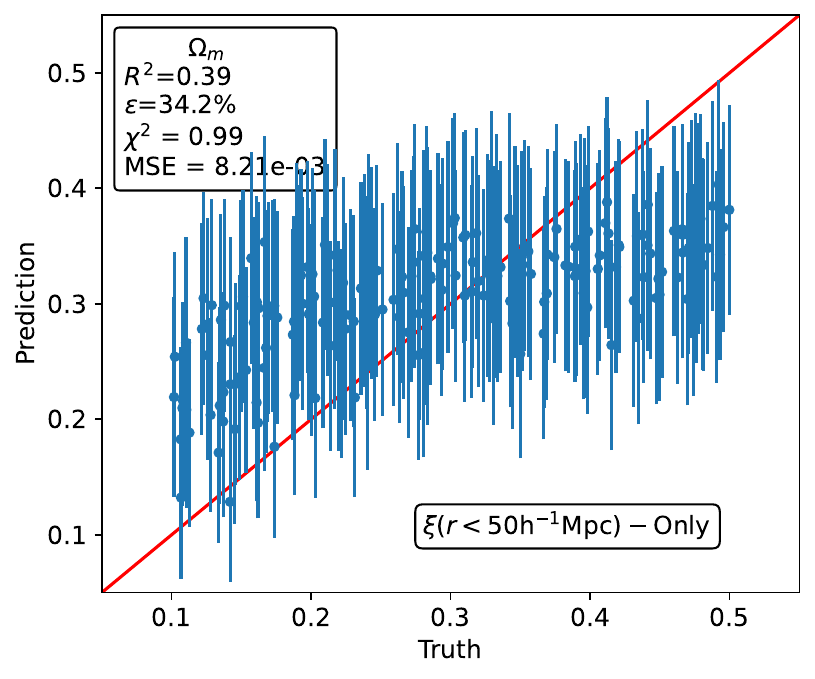}
        \includegraphics[width=0.45\linewidth]{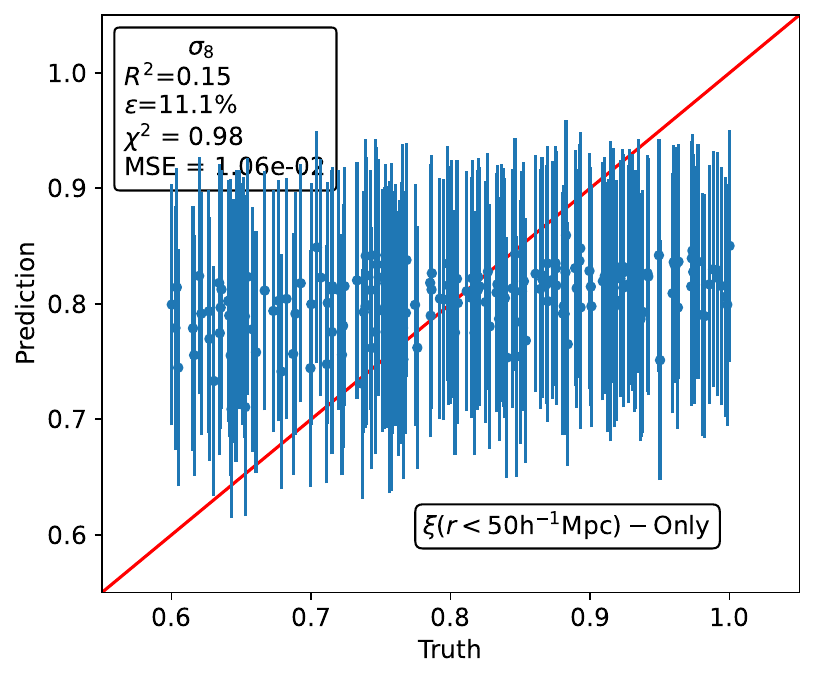}
        \includegraphics[width=0.45\linewidth]{Only_CDF_halo_Om.pdf}
\includegraphics[width=0.45\linewidth]{Only_CDF_halo_Sig.pdf}
    \end{center}
    \caption{Comparison between $\xi(r)$-only and CDF-only. As shown, CDF-only performs much better compared to $\xi(r)$-only, as expected from \citep{Banerjee_2021a}.}
    \label{Fig:2ptCF}
\end{figure*}

As a benchmark, we have demonstrated how the constraining power of the two-point correlation function (2ptCF), $\xi(r)$, compares with the NN CDFs in Fig. \ref{Fig:2ptCF}. We employed Pylians\footnote{\url{https://pylians3.readthedocs.io/en/master/}} to compute the 2ptCF on scales smaller than $50 \, \rm h^{-1}\, Mpc$ (to ensure the same range of scales as for the kNN statistics) on the same halo catalogues of the $10^5$ most massive halos that were used to calculate the kNN statistics. We used these as the input to a fully connected neural network with LeakyReLU activation functions to obtain constraints on the $\Omega_m$ and $\sigma_8$. The number of layers, neurons of the fully connected neural network, along with the weight decay and learning rate of the Adam optimiser \citep{Kingma2014AdamAM} are kept as hyperparameters and optimised using \textsc{OPTUNA}. The results obtained are shown in Fig. \ref{Fig:2ptCF}. As is evident, the performance of the CDFs is significantly better (by almost a factor of two in some of the validation metrics, for example  $R^2$) compared to the 2ptCF for both the parameters. This is also expected since \cite{Banerjee_2021b} found a similar conclusion, although using the Fisher analysis.

\begin{table}
\centering
\caption{Summary of validation metrics for different configurations.}
\label{tab:res_all}

\setlength{\tabcolsep}{3pt}
\renewcommand{\arraystretch}{1.15}

\begin{tabular}{|c |c|c|c|c| c|c|c|c|}
\hline

\makecell{ML\\model} &
\multicolumn{4}{c}{$\Omega_{\rm m}$} &
\multicolumn{4}{c|}{$\sigma_8$} \\

\cline{2-5}
\cline{6-9}

&
$R^2$ &
\makecell{$\epsilon$\\(\%)} &
$\chi^2$ &
\makecell{MSE\\($10^{-3}$)} &

$R^2$ &
\makecell{$\epsilon$\\(\%)} &
$\chi^2$ &
\makecell{MSE\\($10^{-3}$)} \\
\hline

CDF-only &
0.79 & 17.0 & 0.81 & 2.8 &
0.59 & 7.5 & 0.76 & 5.1 \\

Map-only &
0.74 & 17.9 & 0.91 & 3.4 &
0.00 & 13.1 & 1.05 & 13.5 \\

Map+CDF &
0.80 & 15.3 & 1.19 & 2.6 &
0.93 & 3.2 & 0.78 & 0.96 \\
\hline

\end{tabular}
\end{table}

\section{Related works and comparison}

In the following, we compare the performance of our hybrid network with results obtained by the existing models that also utilise halo and/or galaxy positions. While several studies have demonstrated the use of diverse machine learning techniques to constrain cosmological parameters, we limit our comparison to those based on the Quijote dataset as others \citep[e.g.][]{shao_et_al, Pointnet_22} use different datasets, and therefore do not permit a meaningful direct comparison.
    
     In \cite{Ho_SBI_24}, the positions of $\sim 10,000$ halos were used as the input of a GNN to infer the cosmological parameters. In the absence of any values of the accuracy metrics in their paper, we visually estimated (from Fig. 7 in their paper) that their relative error for $\Omega_{m}$ is $ \sim 20\%$, which is poorer compared to our hybrid study. Their estimate of $\sigma_{8}$ is very close to our estimate.

     In \cite{Cuesta_pointcloud_23} the authors used the position of the 5,000 most massive halos as the input features for their generative modelling, and they obtained a mean relative error of $\sim5\%$ and $\sim3\%$ on $\Omega_{\rm m}$ and $\sigma_8$, respectively. While their constraints on $\Omega_m$ are better than our study, the constraints on $\sigma_8$ from our study are comparable with their constraints. Further, caution must be used as the learned likelihood from their model is not well calibrated for $\Omega_{\rm m}$, whereas our study is not affect by any such issues.

     In \citet{Chatterjee_25}, the authors found that a point cloud-based network trained on the position of the 8192 most massive halos constrained $\Omega_m$ with a relative error of $15.5\%$, which is very similar to the results of this study ($15.3\%$). Interestingly, this point cloud-based study (when trained on the position of the halos) failed to infer $\sigma_8$ completely, whereas the present study has been able to recover $\sigma_8$ with excellent accuracy across all the validation metrics. We would like to emphasise that the training time and memory requirements in this study are significantly lower than those reported in \citet{Chatterjee_25}. For instance, training with 8,192 halos and 32 neighbours at a batch size of 32 took approximately 2 days in their study, whereas our current approach completes training in only about 2 hours, which  means it is 24 times more efficient than the point cloud-based method. Moreover, while \citet{Chatterjee_25} required 64 GB of GPU memory for a batch size of 32, our current method requires only a 16 GB GPU. The main reason for this is that we use 2D slices of the NN maps in this study, whereas the point cloud-based method works in three dimensions.
     
     In \cite{Lee_25}, the authors used topological neural networks trained on the positions of the 5,000 most massive halos and recovered $\Omega_{\rm m}$ and $\sigma_8$ with relative errors of $15.33\%$ and $3.69\%$ with their best performing FullTNN network. Both of their constraints are similar to the constraints we find in this analysis.

     In \cite{Cosmobench_25}, the authors used a GNN-based neural network and performed cosmological parameter inference analysis on the Big Sobol Sequence of Quijote simulation suites. When using positions as the only feature for the halos, they obtained an $R^2$ value of $0.80$ and $0.77$ for $\Omega_m$ and $\sigma_8,$ respectively. In our hybrid model, while the $R^2$ value for $\sigma_8$ is better than their reported $R^2$ values, the $R^2$ value for $\Omega_m$ is the same as theirs. Moreover, our hybrid model takes 2 hours of GPU time, whereas their GNN-based model takes 24 hours on TPU.

\section{Discussion and conclusions}
In this study we have demonstrated the use of kNN statistics as inputs to neural network models for inferring cosmological parameters from halo catalogues generated by the Latin Hypercube dataset of the Quijote simulation suite. Specifically, we employed a hybrid neural network architecture that leverages both nearest neighbour distance maps and nearest neighbour cumulative distribution functions (CDFs), successfully recovering $\Omega_m$ and $\sigma_8$ with excellent accuracy.

We conducted a detailed comparison with all existing studies that use the Quijote simulations for cosmological parameter inference and showed that our approach achieves very high accuracy, while remaining highly computationally efficient.

The advantages of our method over previous approaches, such as those based on GNNs, topological neural networks, or point cloud networks, are twofold. First, it provides a far more efficient framework for cosmological parameter inference, substantially reducing computational cost relative to GNN- or PointMLP-based methods. Second, it uniquely enables field-level representations from extremely large halo samples with remarkable efficiency. These features make our model particularly well-suited for forthcoming galaxy surveys, which are expected to produce unprecedentedly large datasets.

One limitation of this study is that NN distance maps or CDFs cannot incorporate the information about the mass and velocity of the halos. In \cite{Chatterjee_25} we find that mass and velocity are very crucial in putting tighter constraints on the cosmological parameters. To overcome this, in future work, we plan to include the peculiar velocities of the halos in one of the spatial directions (using redshift space distortion) in the simulations before making the images. It may even provide us with a stronger correlation between the velocity and the matter field. It is therefore possible for the neural network to pick up the subtle anisotropies being produced by these redshift space distortions, providing even tighter cosmological constraints.

\begin{acknowledgements}
The work of AC was supported by the European Union’s Horizon Europe research and innovation programme under the Marie Skłodowska-Curie Postdoctoral Fellowship HORIZON-MSCA-2023-PF-01, grant agreement No 101151693 (LUPCOS). AB’s work was partially supported by the Startup Research Grant (SRG/2023/000378) from the Science and Engineering Research Board (SERB), India. This work was also supported by U.S. Department of Energy grant DE-AC02-76SF00515 to SLAC National Accelerator Laboratory managed by Stanford University. The work of FVN is supported by the Simons Foundation. The authors acknowledge the PARAM Brahma Facility under the National Supercomputing Mission, Government of India, at the Indian Institute of Science Education and Research, Pune, for providing the computing resources for this work. The ML architecture developed in this work is implemented in \textsc{PyTorch} \citep{paszke2019pytorch}. The authors acknowledge the use of CHATGPT for refining the text at the final stage of the manuscript.
\end{acknowledgements}

\bibliographystyle{bibtex/aa}
\bibliography{main} 
\begin{appendix}

\section{Constraints on cosmological parameters from dark matter particles}
\label{sec:DM_constraints}
\begin{figure*}
\begin{center}
\includegraphics[width=0.45\linewidth]{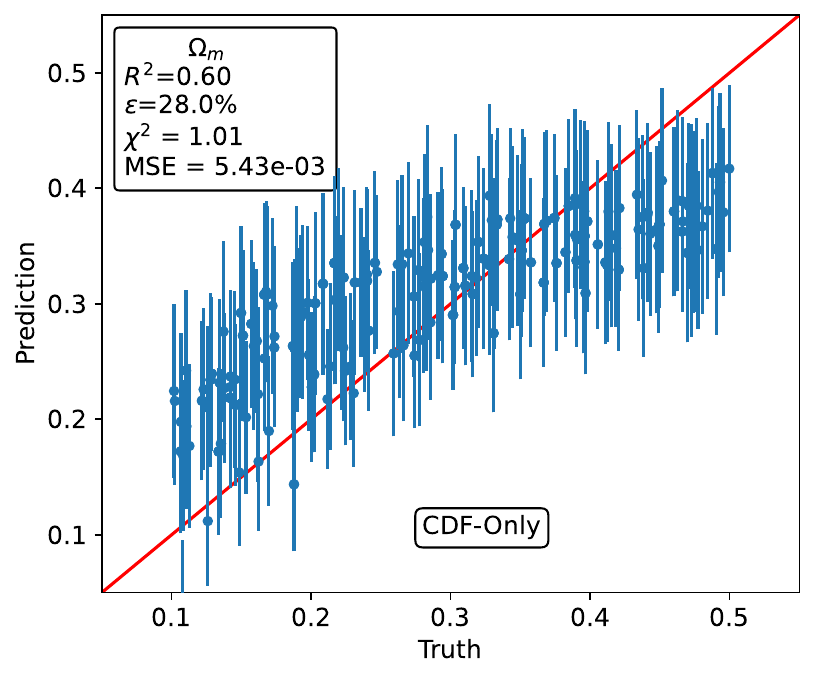}
\includegraphics[width=0.45\linewidth]{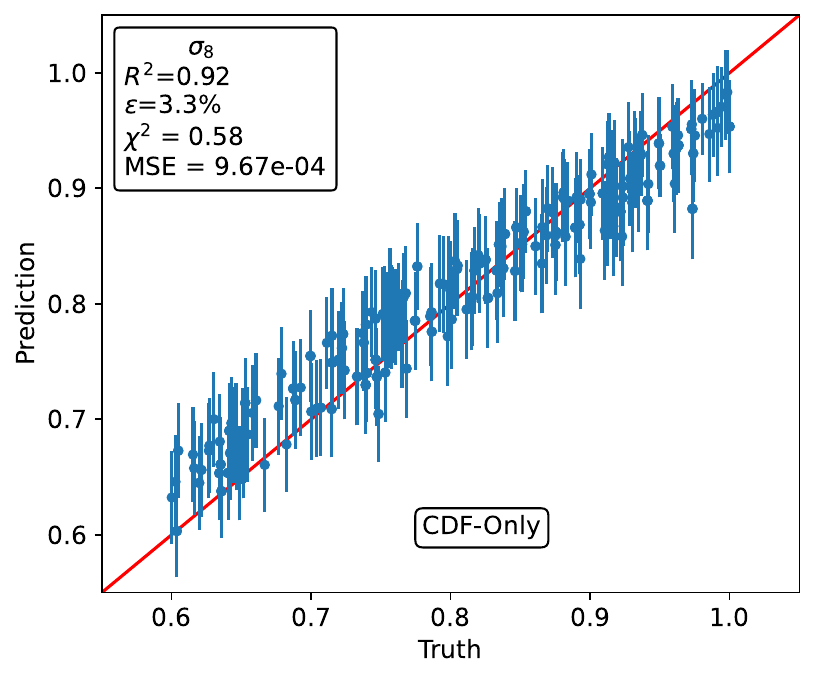}
\includegraphics[width=0.45\linewidth]{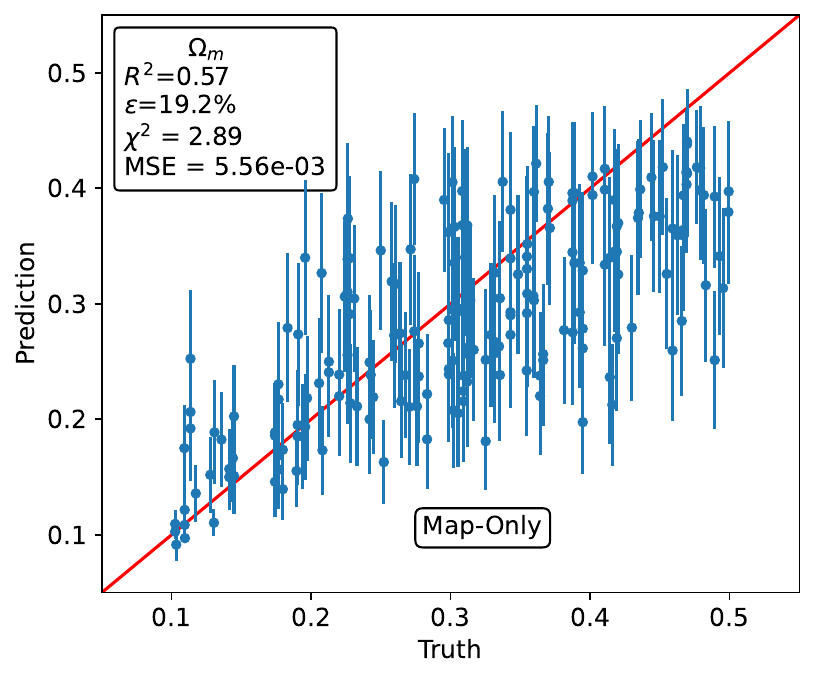}
\includegraphics[width=0.45\linewidth]{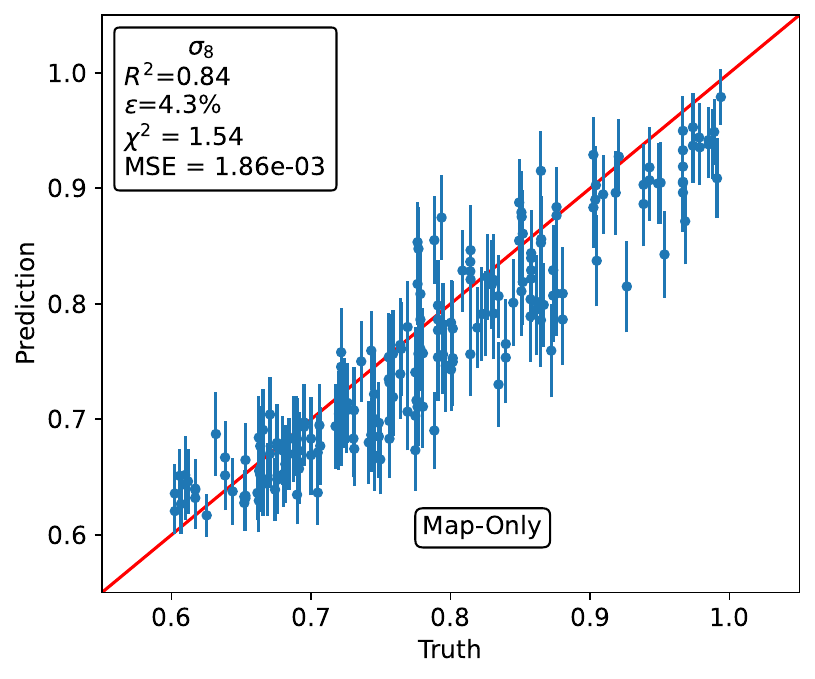}
\includegraphics[width=0.45\linewidth]{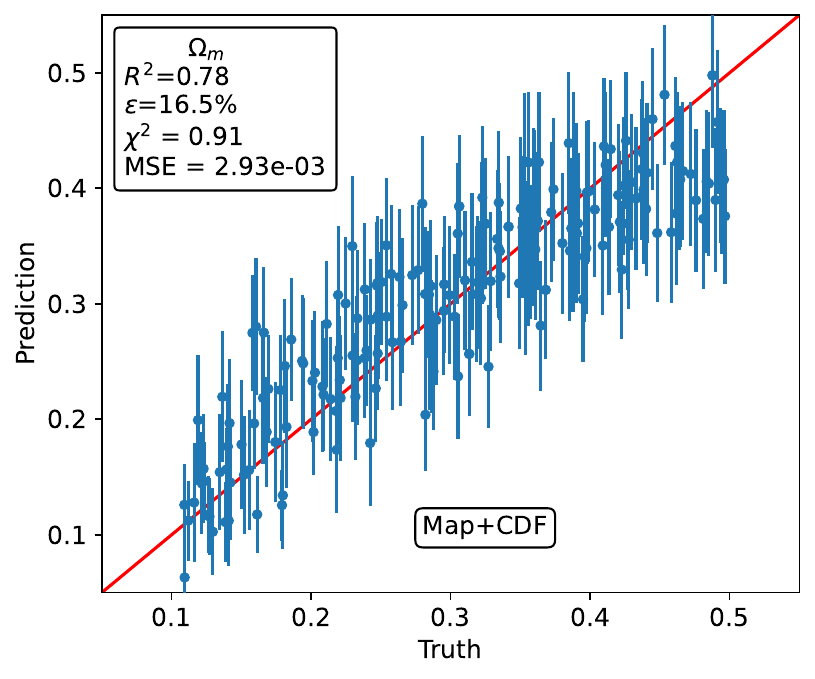}
\includegraphics[width=0.45\linewidth]{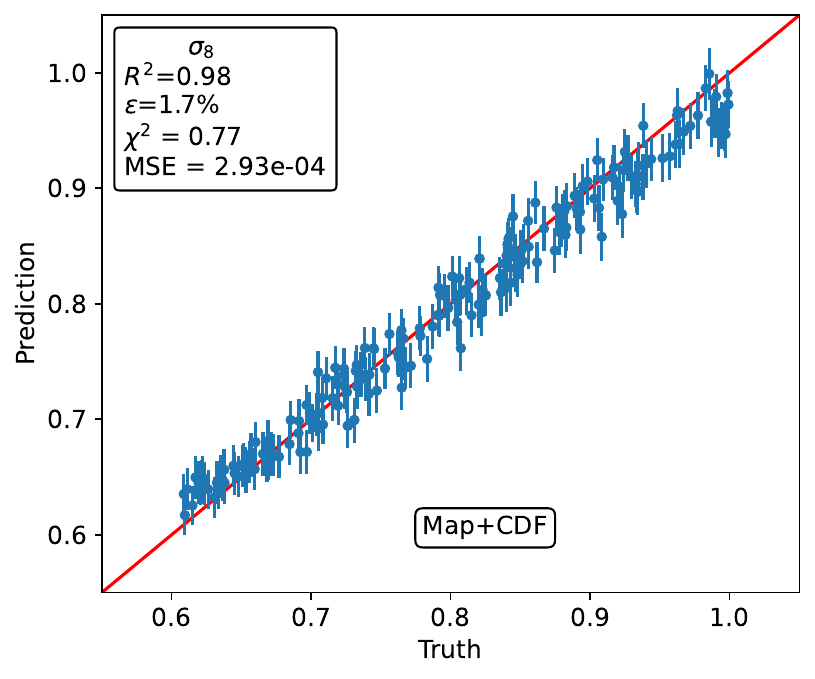}
\caption{Same as Fig \ref{fig:Om_sig8} but for DM particles rather than halo catalogeus}
\label{fig:Om_sig8_DM}
\end{center}
\end{figure*}

As a proof of concept, we compute NN maps and NN CDFs for dark matter (DM) particles to test their constraining power on cosmological parameters, both individually and in combination. In this analysis, we follow the same procedures outlined in Sects. \ref{subsec:NN_CDF} and \ref{subsec:NN_maps}, but instead of DM halos, we use the three-dimensional positions of $10^5$
DM particles. The resulting constraints are shown in Fig. \ref{fig:Om_sig8_DM}. We find that, while the Map-only model outperforms the CDF-only model for $\Omega_m$, the situation is reversed for  $\sigma_8$. When the Maps and CDF are combined, i.e. Map+CDF, their joint constraining power exceeds that of either method alone.

\begin{table}
\centering
\caption{Final values of the hyperparameters obtained using OPTUNA for different ML models.}
\label{tab:hyperparameters_all}

\setlength{\tabcolsep}{4pt}
\renewcommand{\arraystretch}{1.15}

\begin{tabular}{|c| c|c|c|c|c|}
\hline

\makecell{ML\\model} &
\multicolumn{5}{c|}{Values of Hyperparameters} \\
\cline{2-6}

&
\makecell{Learning\\rate} &
\makecell{Weight\\decay} &
\makecell{No. of\\layers} &
\makecell{Dropout\\rate} &
\makecell{Neurons in\\each layer} \\
\hline

CDF-only &
0.0017 &
$12.8 \times 10^{-5}$ &
1 &
0.21 &
400 \\
\hline

Map+CDF &
0.0074 &
$1.21 \times 10^{-8}$ &
2 &
0.6, 0.3 &
45, 184 \\
\hline

\end{tabular}
\end{table}

\end{appendix}
\end{document}